# An LHCb Vertex Locator (VELO) for 2030s

Evangelos Leonidas Gkougkousis[1], on behalf of the LHCb VELO group

[1]*European Organization for Nuclear Research - CERN, Espl. des Particules 1, 1211 Meyrin, Switzerland*

*E-mail:* egkougko@cern.ch

The Upgrade II of the LHCb detector, foreseen for 2031, will operate at an instantaneous luminosity of $1.5 \times 10^{34}$ cm$^{-2}$s$^{-1}$, accumulating a sample of more than 300 fb$^{-1}$. To cope with the estimated pile-up of 42 and 200 charged particle tracks per event, precise timing will be added to the tracking and vertexing sub-systems. A new Vertex Locator (VELO), capable of managing the expected 7.5-fold increase in data rate, occupancy, and radiation fluence is needed. Based on a 4D hybrid silicon pixel technology, with enhanced rate and timing capabilities on the ASIC, the new VELO will allow for precise *beauty* & *charm* hadron identification and real time pattern recognition. Through detailed simulations, the fluence, inner radius, material budget and pixel size phase space are explored, while constraining the Impact Parameter (IP) resolution to the Upgrade I value. Two distinct scenarios emerge as starting points for further optimizations, with inner radii and end of life fluence of 5.1 mm at $6 \times 10^{16}$ n$_{eq}$/cm$^2$ and 12.5 mm at $8 \times 10^{15}$ n$_{eq}$/cm$^2$ respectively. Advances and current R&D on sensor technologies, including LGADs, 3Ds and planar pixels are reviewed, focusing on radiation hard designs and defect engineering. ASIC related requirements with respect to sensor capacitance and power budget are taken into consideration for achieving the 30 ps per hit timing target towards the future 28 nm protype submission. Improvements in cooling, mechanics and vacuum implementations are examined with respect to each layout scenario. The use of bi-phasic Krypton cooling is evaluated as an option for the case of above 1.5 W/cm$^2$ power dissipation. Replaceable sensor modules, coupled with 3D printed titanium supports are also under consideration. Finally, a comprehensive R&D schedule towards final design optimization within a six-year period is discussed.

## 1. Introduction

The LHCb experiment at CERN is a general-purpose forward spectrometer optimized for heavy flavour physics and rare decays. It provides a coverage of 2 < η < 5 with a momentum resolution of 0.5% at $p_T$ < 20 GeV. A dataset of 10 fb$^{-1}$ has been collected at the end of Run 2, followed by a major detector upgrade (Upgrade I), fully replacing the vertexing and triggering subsystems [1]. With an estimated dataset of 50 fb$^{-1}$ at the end of Run 4, a second upgrade (Upgrade II) is planned, to be implemented for Run 5 & 6 operation (2035 onwards). The LHCb Upgrade II will operate at instantaneous luminosities of the order of $15 \times 10^{33}$ cm$^{-2}$s$^{-1}$ towards the 300 fb$^{-1}$ mark (see Fig. 1), with an average PileUp of μ = 42 visible interactions per bunch-crossing and an estimated charged track multiplicity of 200 [2]. Interaction vertices will be distributed following a gaussian profile over a 186 ps time window, with a 45 mm sigma along the z-axis (see Fig. 2). To cope with this 7.5-fold increase in luminosity and track density with respect to Upgrade I, major upgrades to the experiment are necessary. During Upgrade II, a large area pixel detector within the dipole magnet will be added, coupled with a time-of-flight station (TORCH). The latter will offer a 70 ps per hit time resolution enhancing low momentum particle reconstruction and identification [3]. Timing information will also be added to the current RICH and calorimeter detectors. Most importantly, a new vertexing subsystem providing 20 ps resolution per track is planned while maintaining or improving current position resolution.

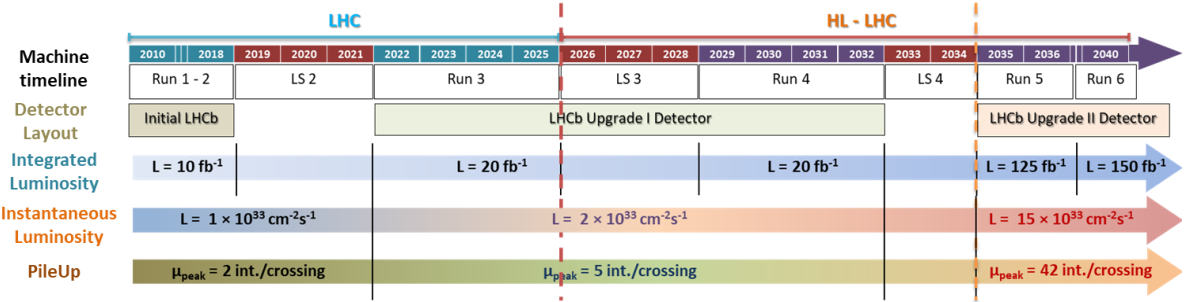

**Fig. 1.** Latest LHC schedule. Vertical lines mark the evolution of the LHCb detector from the original layout through Upgrade I & Upgrade II, towards the final goal of the 300 fb$^{-1}$ dataset.

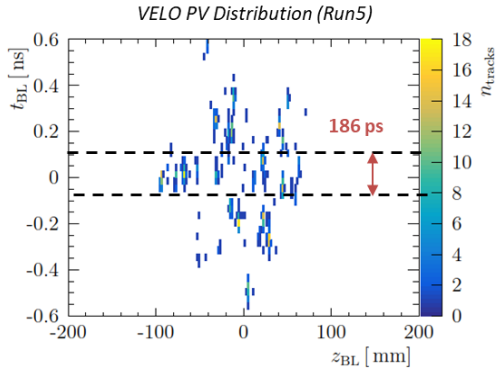
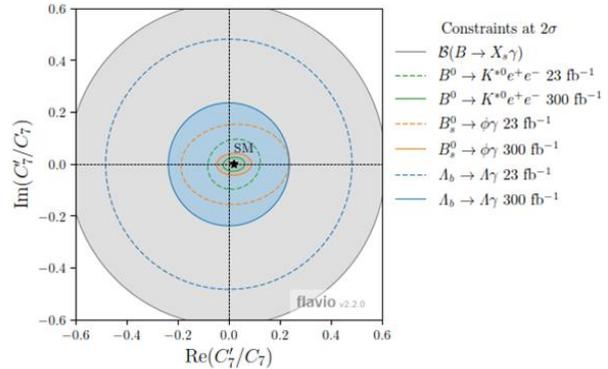

**Fig. 2.** Spatial and temporal distribution of primary vertices during Run 5. The 186 width (2σ) Gaussian window is marked (dashed lines).

**Fig. 3.** Projected Upgrade II sensitivity to the real and imaginary part of the ratio of right- and left-handed Wilson coefficients, $C'_7$ and $C_7$, compared to the expected Run 3 sensitivity [3].

LHCb Upgrade II is designed to fully exploit the flavour-physics opportunities of the HL-LHC. Since discrepancies in the side measurements of the CKM matrix unitarity triangle can be indicators of new physics, precise measurement of the angle γ is of particular interest. While currently known to a 5° precision, LHCb Upgrade II will allow for an order of magnitude improvement, with an expected 0.35° uncertainty at the end of Run 6 [4]. In addition, for branching ratio measurements in flavour changing transitions, notably the B($B^0$→μ$^+$μ$^-$)/B($B^0_s$→μ$^+$μ$^-$) ratio - exhibiting 1.5 σ level anomalies with respect to Standard Model (SM) predictions [5] - Upgrade II statistics will increase the precision from 34% to 10% (see Fig. 3), expanding angular distributions analyses [6].

## 2. VErtex LOcator (VELO) for Upgrade II

The Vertex Locator (VELO) is a hybrid silicon detector asymmetrically spanning both directions of the z-axis around the interaction point [1]. Tasked with precise primary and secondary vertex reconstruction, it operates in a high rate and track multiplicity environment. Under Run 5 conditions, mean primary vertex (PV) separation evolves from the current 4.25 mm to 1.5 mm, which is inferior of the PV z-position uncertainty at forward η (see Fig. 4). Maintaining current position resolution during Upgrade II would therefore be insufficient and a decrease in efficiency to the level of 60% is to be expected. Performance can be recovered by attaching a 20 ps time resolution to each track, as demonstrated in Fig. 5, for events including at least one $b$-hadron.

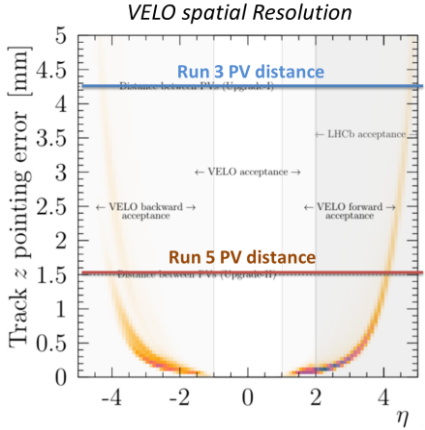
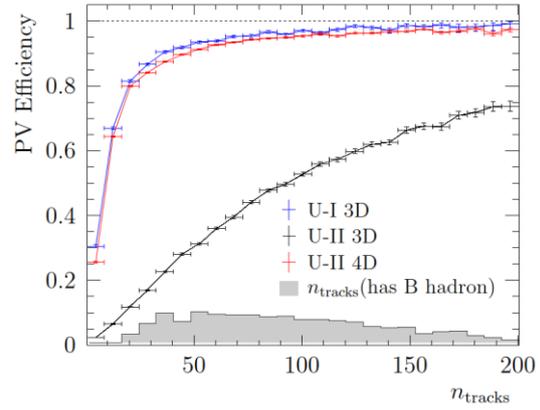

**Fig. 4.** Track position uncertainty versus η from MC simulation. VELO acceptance and mean PV distance for Runs 3 & 5 are denoted.

**Fig. 5.** Primary vertex efficiency for Upgrade II with (red) and without (black) timing. Current Upgrade I performance (blue) noted as a reference.

LHCb's triggerless scheme, with online event selection for most $b$-related decay channels, allows for minimal recourse once a decision propagated. Correct identification of displaced vertices for long flying $b$-hadrons is therefore essential for ghost track rejection and background suppression. Association of secondary $b$-vertices with only two tracks to the origin PV becomes challenging in a high multiplicity forward tracking environment. While timing primarily serves to recover PV separation, a per-track timestamp would allow correct association of secondary vertices to the PV whose timestamp they share. Simulated studies with a simplified Kalman filter approach at $B^0_s \rightarrow D^-_s \pi^+$ events demonstrate an efficiency recovery of up to 90% for an IP > 0.1 mm with the addition of timing [7].

Several approaches for timing implementation have been studied, within the 20 ps per track resolution envelope. Combinations of a vertexing system with dedicated timing planes, primarily at the endcap regions, require a minimum of three such layers for efficient extrapolation and outlier rejection. Three such variations were considered, including small / large timing planes at high η regions (2.8 / 2 < |η| < 5) and barrel

stations (see Fig. 6). While total timing silicon area in such options is limited with respect to a full-fledged 4D-vertex detector, single hit resolution is reduced from 50 ps to 25 ps. Furthermore, an estimated pixel size of 100 x 100 μm$^2$ would be needed to keep occupancy within the 10% level. Additional complications arise by the need for two dedicated ASICs and the increased computing power for reconstruction, due to the absence of efficient pattern recognition and lower PV efficiency (see Fig. 7). As a result, a complete 4D option is privileged, where timing and position information originate at the same layer.

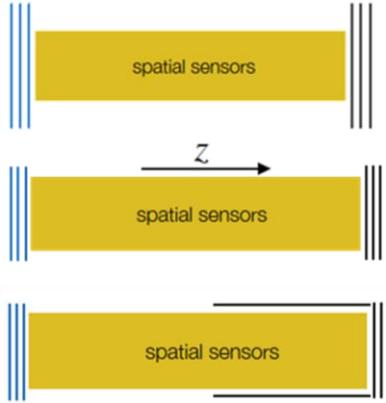 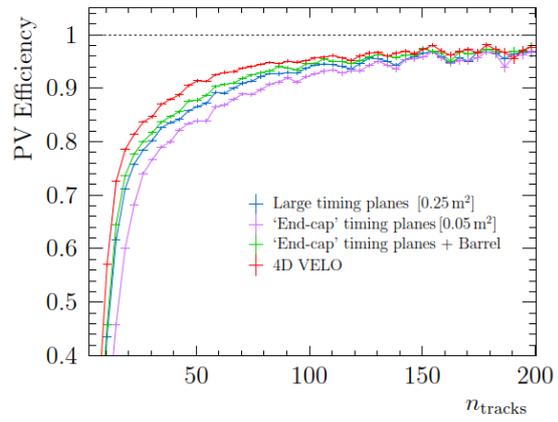

**Fig. 6.** Timing layout variations, including large/small endcap timing planes (top/middle) and endcap plus barrel timing planes (bottom).

**Fig. 7.** PV reconstruction efficiency for each timing layout variation.

An ideal 4D VELO layout would minimize radiation damage while maintaining IP position resolution at Upgrade I levels. The latter, can be expressed as a convolution of an extrapolation term, related to single hit resolution, and a scattering term, connected to the $X_0$ of sensors and support structures. The extrapolation term depends on intrinsic sensor resolution and distance from the interaction point. Similarly, because of the vertical orientation of the modules with respect to the beam plane, radiation fluence can be factorized with respect to inner radius as $\Phi = \alpha \times R^b$, where b depends on module position along the z-axis, varying between -1.8 to -2.3 [3]. Consequently, three main parameters are to be considered during detector layout optimization: $X^0$, pixel size and inner radius (see Fig. 8).

**Table I.** Characteristics of scenarios considered as starting point for VELO Upgrade II layout optimization.

| Parameter | Scenario A ($S_A$) | Scenario B ($S_B$) |
|---|---|---|
| Inner Radius | 5.1 mm | 12.5 mm |
| Hit Rate (hottest pixel) | 350 kHz | 40 kHz |
| Bandwidth | 125 (Gb/s)/cm$^2$ | 47 (Gb/s)/cm$^2$ |
| Pixel Size | 55 × 55 μm$^2$ | 42 × 42 μm$^2$ |
| EOL fluence | $6 \times 10^{16}$ n$_{eq}$/cm$^2$ | $8 \times 10^{15}$ n$_{eq}$/cm$^2$ |
| $X_0$ requirements | same as Upgrade I | 1/5 × Upgrade I |

Exploiting the above optimization scheme, two extreme scenarios are considered as starting point for further layout consolidation (Table 1):

- ✓ **Scenario A:** Current inner radius of 5.1 mm is maintained, leading to an estimated 7.5-fold increase in hit rate and a 6-fold increase of radiation fluence with respect to Upgrade I. Nevertheless, relaxed $X_0$ constraints allow to maintain the current RF foil geometry and detector layout, but extreme radiation hard sensors and ASIC technologies will need to be developed. To mitigate the issue, possible periodic replacement of modules can be envisaged.
- ✓ **Scenario B:** Sensors are placed further away from the interaction point with an inner radius of 12.5 mm, keeping radiation fluence and hit rate at Upgrade I levels. In contrast, to compensate for the extrapolation term degradation, spatial resolution will need be improved to 9 μm with a drastically reduced material budget. Thickness reduction on ASIC, sensors and RF-foil will have to be implemented, leading to a complete redesign of the current geometry. To maintain the acceptance, the detector layout will need readjustment with possible addition of sensor planes at both ends, increasing the total length.

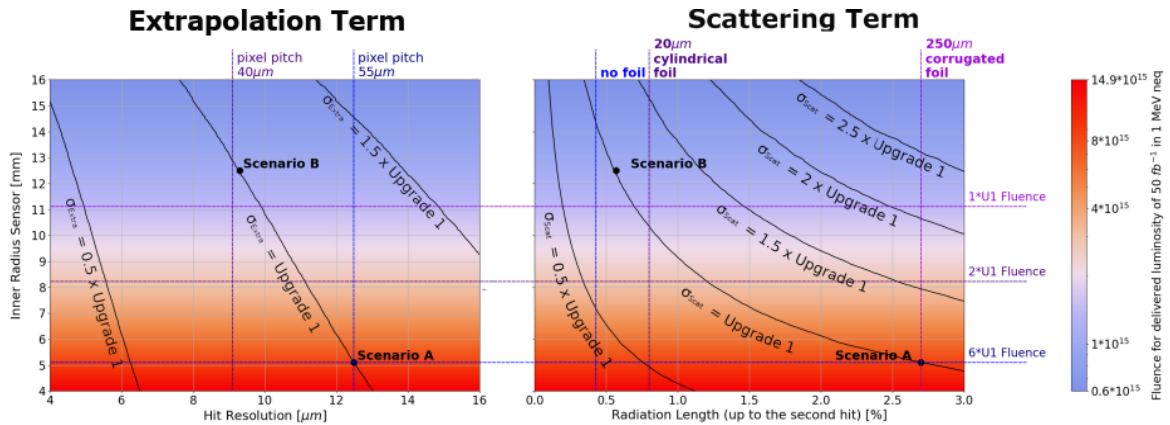

**Fig. 8.** Hit resolution (left) and radiation length (right) versus fluence and inner radius. Iso-resolution working point curves are added referring to Upgrade I IP position resolution.

## 3. Subsystems R&D

### 3.1 Sensor Technologies

Hybrid silicon sensors are currently a more technologically mature approach with respect to the strict timing, radiation hardness and power dissipation requirements of the VELO Upgrade II. While traditional thin planar pixels (50/100 μm) can be considered, the amount of deposited charge (3.6 and 7.2 ke respectively before irradiation, with an estimated 60% reduction at fluences ~$1 \times 10^{16}$ $n_{eq}/cm^2$ [8]) makes it challenging for the electronics to deliver the required time resolution within the allocated power budget. Alternative options include, but are not limited to, 3D sensors, Low Gain Avalanche Detectors (LGADs) and several monolithic implementations. All these technologies exhibit excellent timing characteristics.

LGADs offer similar drift distances as thin planars (50 μm thickness), resulting in fast signals, but generate significantly more charge due to gain of 20 – 40, achieved through a secondary implant (gain layer). These implants are extremely sensitive to radiation, with complete deactivation (acceptor removal) at fluences above $5 \times 10^{15}$ $n_{eq}/cm^2$. Although the end-cap timing layers of ATLAS (HGTD) [9] and CMS (HG-CAL) [10] have opted for this technology, radiation hardness requirements in these applications are limited to $2 \times 10^{15}$ $n_{eq}/cm^2$ and $1 \times 10^{15}$ $n_{eq}/cm^2$ respectively. Towards a more radiation hard LGAD design, different gain layer implementations were studied. A comparison between standard boron, gallium, and boron with deeply diffused carbon gain layer implementations (see Fig. 9) demonstrated a 20% improvement by the addition of carbon in the bulk. In contrast, substitution of boron with gallium resulted to a 20% reduction in radiation tolerance [11]. The possibility of lithium co-implantation and indium gain layer to increase radiation hardness by mitigating acceptor removal via defect engineering is currently under investigation [12].

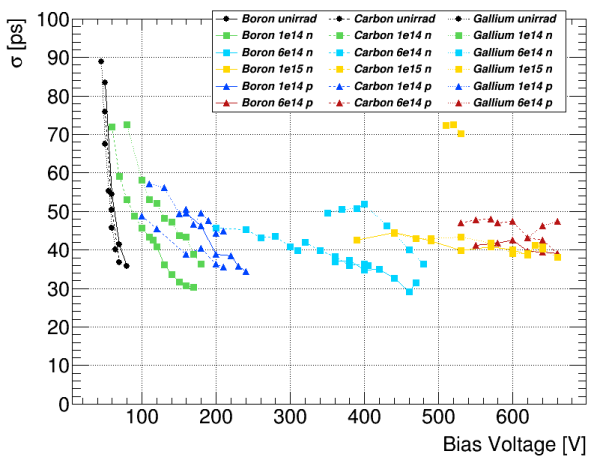
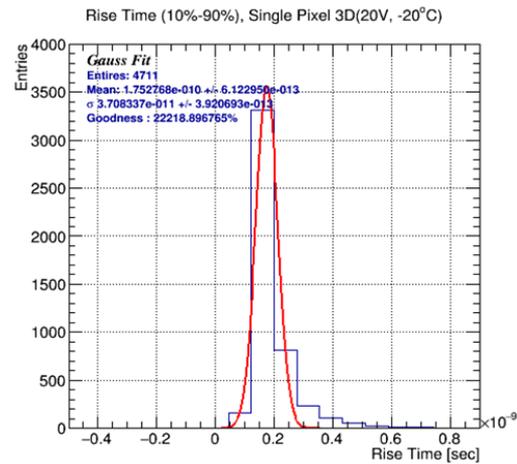

**Fig. 9.** Time resolution vs bias voltage of gallium, boron, and boron + carbon diffused gain layer proton and neutron irradiated LGADs at -30 $^0$C using MIPS [11].

**Fig. 10.** MIPS induced signal rise time distribution of a 55 x 55 μm$^2$, non-irradiated, single 3D pixel at -20 $^0$C [13].

3D silicon pixel geometries present several advantages by decoupling the drift and charge generating directions. Such sensors combine extremely fast signals with rise time in the order of 180 ps (see Fig. 10), with increased collected charge and exceptional radiation hardness ($> 3 \times 10^{16}$ $n_{eq}/cm^2$) [14]. While traditional column geometries suffer from field non uniformities introducing jitter, trench-based approaches mitigate the effect (TIMESPOT) [15]. Time resolution of the order of 12 ps have been achieved while column-based geometries present the appealing possibility of charge multiplication at the collection electrode, mostly after high irradiation, due to the increased bias voltage. Extensive test-beam campaigns with irradiated structures of both types to the highest expect levels of Upgrade II are under way to evaluate these technologies [13].

Several monolithic approaches have been proven to yield excellent time resolution in recent years (DMAPs, SiGe BiCMOS [16]). Lack of sufficient radiation hardness and of high readout rate capabilities on the digital part though, make such technologies ill-adapted for VELO Upgrade II at the current stage.

*3.2 ASIC design*

Design of a suitable ASIC, within the time resolution, power budget, radiation hardness, bandwidth, and footprint requirements, is extremely challenging. In an ideal charge amplifier, hard physical limits on achievable time resolution with respect to input capacitance and charge exist (see Fig. 11) [17]. These quantities are directly related to the pixel size and active thickness. At a 130 μm active thickness, yielding 10 ke for pad capacitance of 110 fF (TIMESPOT geometry) [18] one could expect > 20 ps per hit within the 1.5 W/cm$^2$ power density envelope. An increase in pixel size and/or decrease of active thickness would result in an elevated input capacitance, and in term affect the jitter (see Fig. 12). Similarly, collected charge depends not only on active thickness (~0.010 fQ/μm for non-irradiated silicon), but also on incidence angle and active volume, affecting time jitter. These limitations are considered on final sensor design and impose hard limits on feasible geometries.

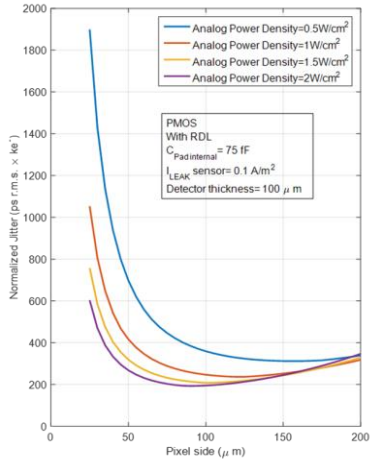 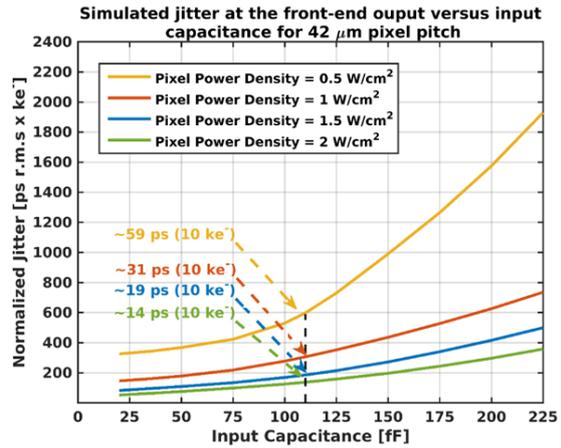

**Fig. 11.** Estimated normalized Jitter with respect to input capacitance, expressed as a function of the pixel size for different power densities [17].

**Fig. 12.** Normalized jitter vs input capacitance for different power density scenarios. Jitter is expressed per ke of collected charge [17].

Constraints are also imposed with respect to pixel footprint, as the available area for integration varies with the square of the pitch. Though the choice of 28 nm CMOS technology has been made, with blocks tested for radiation hardness up to 10 MGy [19], the analog part's footprint is not expected to significantly shrink with respect to current 35 nm integrated ASICs (Timepix4, VELOPiX) [20], occupying 25% of available area for a 55 μm pixel pitch. As a result, available footprint for the digital part shrinks by a factor of 2.3 between the two scenarios (42 μm vs 55 μm pitch), passing from ~2270 μm$^2$ to ~1000 μm$^2$. The target performance of small-scale prototype towards a full Upgrade II ASIC (PicoPix), based on 28 nm integration, is < 30 ps time resolution at 1.5 W/cm$^2$ power consumption with a data rate of 125 Gbps/cm$^2$. An initial submission is expected within the following 18 months.

*3.3 Wake-field manifold*

The proximity of sensor modules to the beam envelope (5.1 mm distance), within the secondary vacuum, in conjunction with high beam currents (1.1 A [21]), require adequate management and propagation of beam-induced wake-fields along the length of the detector. A corrugated aluminium foil of 180 μm mean thickness is currently serving this purpose, also acting as a barrier between the primary and secondary vacuum. Though a final decision towards an Upgrade II implementation has not yet been reached, three different options are proposed:

1. A 20 μm thick cylindrical aluminum foil, spanning the length of the VELO and surrounding the beam envelope. The structure will be tensioned at both edges for stability. NEG (Non-Evaporative Getter) coating commonly applied to limit outgassing and vacuum contamination can in this case be a large part of estimated $X_0$ due to heavy metal presence (Ti, Zr, Hf, V) [22]. Alternatives, including thin amorphous carbon coating (0.4 μm), are under investigation.
2. A mesh of 70 μm diameter wires at 200 μm pitch surrounding the beam envelope, exhibiting similar electrical properties as a 19 μm aluminum foil.
3. A carbon composite structure. Although estimated $X_0$ can be as high as 1.5 times that of an aluminum foil, dual use of the structure as module support frame can ultimately lead to material budget reduction. This option is more convenient for smaller thickness (< 200 μm) and complex shapes.

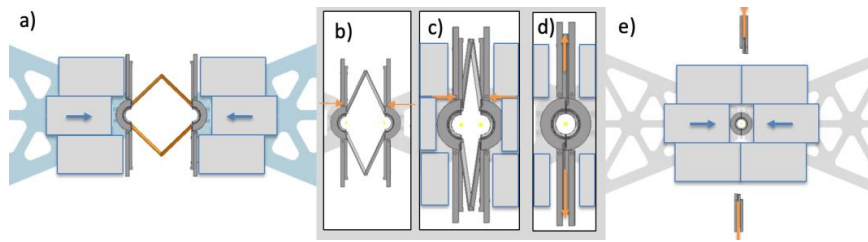

**Fig. 13.** Possible implementation of a non-fixed wake-field manifold, for a VELO implementation within the primary vacuum. Foldable conductive structures expand to ensure electric isolation between the modules and the beam while the velo is retracted (yellow surfaces at a). As the VELO approaches its closed position (b, c and d), the surfaces fold towards the walls of the fixed part of the wake-field manifold, until they become completely flush with the structure (position e).

Final geometry will heavily be impacted by the decision to maintain the VELO at the secondary or displace it within the primary vacuum. To ensure vacuum separation in the latter case, while allowing for the detector to retract during the injection phase, movable separators may be implemented as part of the structure (see Fig. 13).

*3.4 Cooling*

Evaporative $CO_2$ cooling is currently used to mitigate reverse annealing of silicon radiation damage effects and manage front-end power dissipation, allowing for operation down to -40 °C. While power budget is expected to remain at the present levels of 1.5 W/cm$^2$, alternatives, such as bi-phasic Krypton cooling, are considered. An excellent candidate due to its radiation hardness and low melting temperature at its triple point (-157.24 °C at 0.7315 bar [23]), its use requires dedicated R&D to develop appropriate cooling plant components.

Several approaches for heat extraction are investigated, including microchannel plates in direct contact with the modules (see Fig. 14), as implemented for Upgrade I. The high thermal conductivity (150 W/mK) [24], low material budget and a matching Thermal Expansion Coefficient with the sensor modules constitute such an approach an ideal solution. However, the reliance of a microchannel solution to large area direct bonding reduces yield and is prone to defects. Development of "buried-channels" within the wafer, sealed by poly-Si, Si-N or $SiO_2$ or the use of thermocompression with an intermediate metal/ceramic layer are promising alternatives [25]. Additive manufacturing techniques are a more conservative approach. Prototypes have been manufactured using Grade 2, 3D printed titanium (see Fig. 15). Such structures are leak tight for wall thickness in excess of 250 μm, withstanding pressures up to 250 bar. SiC, despite relatively lower thermal performance (120 W/mK), is also a candidate due to its lower cost, higher robustness, lower material budget and proven industrial experience, which are significant advantages.

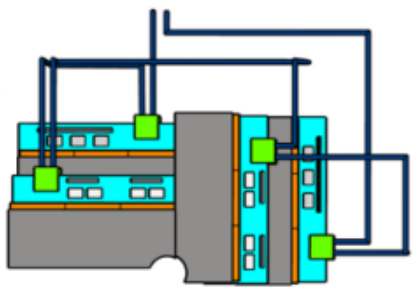
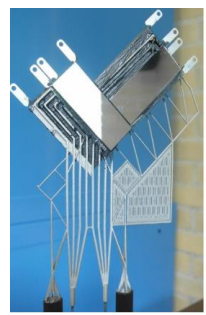

**Fig. 14.** Micro-channel plate implementation design with colling flowing serially between modules.

**Fig. 15.** Prototype 3D printed Titanium cooling plate and support structure, as part of the Upgrade I R&D phase.

*3.5 Vacuum Considerations*

A redesign of the current vacuum tank would be essential for Upgrade II operation, with a geometry mainly dictated by detector layout decisions. The structure must provide vacuum separation, allow for VELO movement, enable periodic replacement if needed, integrate cooling feedthroughs, and support for all beam pipe elements (wake-field manifold).

The intense material budget reduction that a scenario B approach entails, renders the separation between primary and secondary vacuum impractical. In this case, NEG coated surfaces will have to be baked-out and materials carefully chosen to control outgassing while, a dedicated window allowing for this process to complete during shut-downs will have to be implemented. A piston geometry (see Fig. 16) might be a viable solution, where a co-central tubular structure slides along the beam axis sealing the primary vacuum. The detector cassettes can then be accessed by either side for replacement and/or adjustment.

## 4. Summary

Unlocking the full potential for heavy flavour physics at the LH-LHC era entails operating in a harsh environment in terms of PileUp (42), radiation damage ($6 \times 10^{16}$ $n_{eq}$/cm$^2$) and track multiplicity (200 tracks/event). Maintaining current performance is achievable by introducing a 20 ps per track time resolution. Two points in the radiation damage, position resolution and material budget phase space emerge as limits for final VELO Upgrade II optimization, with final layout contained in the phase-space enclosed within these two scenarios:

1. Keeping position resolution and material budget identical to Upgrade I values. Radiation hardness and data rates though would experience a 6- and 7.5-fold increase respectively.
2. Maintaining the same radiation damage and data rate levels with respect to Upgrade I. Position resolution would however have to be improved by reducing the pixel pitch from 55 μm to 42 μm while material budget before the first hit requires a drastic reduction up to 75% in comparison with current implementation.

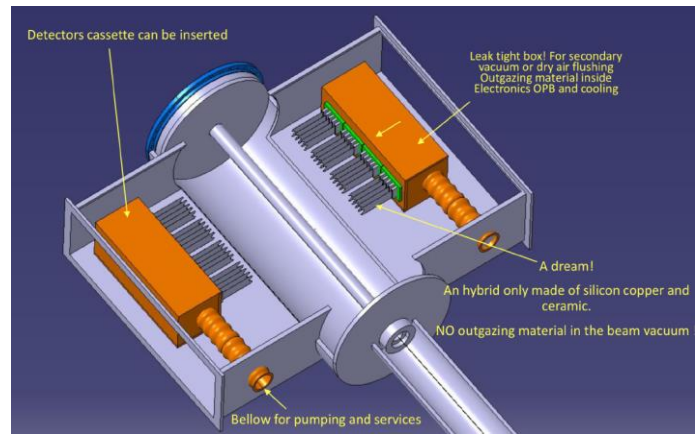

**Fig. 16.** Possible re-design of vacuum tank with no primary vacuum separation and wire-mesh wake-field shield. Central piston depicted in closed position, allowing for side extraction of detector cassettes.

Intensive R&D on sensor technologies (including LGADs, 3Ds and planar pixels) and ASIC designs (28 nm integration for up to 10 MGy) has been undertaken, exploring the limits of available technologies. A layout decision, based on the conclusion of such feasibility studies, will be reached within a 2-to-3-year window. Further improvements on the vacuum, wake-field manifold and cooling will then be implemented at that stage, leading towards a final complete design for Upgrade II VELO in a 5-to-6-year horizon.